\documentclass[aps,prl,twocolumn, superscriptaddress]{revtex4-2}  
\usepackage{graphicx}  
\usepackage{dcolumn}   
\usepackage{bm}        
\usepackage{amssymb}   

\usepackage{physics}
\usepackage{xcolor}
\usepackage[normalem]{ulem}

\begin{document}

\preprint{APS/123-QED}

\title{Preparation and coherent manipulation of toroidal moments in molecules}

\author{Kieran Hymas}
\affiliation{Commonwealth Scientific and Industrial Research Organisation (CSIRO), Clayton, Victoria 3168, Australia}

\author{Alessandro Soncini}%
\email{alessandro.soncini@unipd.it}
\affiliation{Department of Chemical Sciences, University of Padova, Via Marzolo 1, 35131 Padova, Italy}

\date{\today}

\begin{abstract}
Molecules with an odd number of electrons typically display paramagnetic behaviour in a uniform magnetic field. Single-molecule toroics---a family of open shell lanthanide complexes---instead display an unprecedented vanishing magnetization. The anomaly is reconciled by degenerate quantum states where electron spins and orbital currents give rise to time-odd and space-odd magnetic vortices known as toroidal moments, that carry a vanishing magnetic dipole. Resilient to stray magnetic fields and susceptible to electric manipulation, toroidal moments have attracted interest for spintronic, magnonic, and photonic applications. While macroscopic toroidal moments feature in some extended systems, molecular toroidal states have yet to be observed, as it remains unclear how existing experimental set-ups could split degenerate states carrying counter-rotating vortices. We propose a realistic pulsed radiation protocol to polarise and observe molecular toroidal moments in a class of MDy$_6$ (M = Al$^{3+}$, Cr$^{3+}$) molecules with coupled Dy$_3$ toroidal moieties. Three resonant MW-pulses---delivered sequentially or simultaneously---selectively and coherently transfers population to a long-lived toroidally-polarised state whose ensuing magneto-electric properties provide a read-out mechanism. Our results provide a strategy to measure and coherently manipulate toroidal states in molecular systems, which is expected to trigger applications of molecular toroidal states to quantum technologies.
\end{abstract}

\maketitle


Atomic spins and orbital currents in molecules can give rise to a magnetic vortex configuration known as a toroidal (or anapole) moment $\bm{\tau}$, equivalently described either as a current winding up a toroidal surface (hence the name \emph{toroidal moment}), or as a circular head-to-tail vortex arrangement of magnetic dipoles, leading to the formal absence of magnetic poles (hence the name \emph{anapole moment}). The magnitude and spatial orientation of this magnetic vortex are described by a polar vector $\bm{\tau}$ oriented normal to the vortex plane, as first described in the context of nuclear physics by Zeldovich in 1957~\cite{zeldovich1957}. According to the choice of gauge, $\bm{\tau}$ can be rigorously defined either as the antisymmetric part of a magnetic quadrupole~\cite{faglioni2004}, or as the lowest moment (toroidal dipole) of an independent toroidal multipole series complementary to the usual magnetic and electric multipolar expansions~\cite{spaldin2008toroidal}. Both descriptions are equivalent and are related by a gauge transformation. Despite lacking an overall magnetic moment, a pure toroidal moment is fundamentally of magnetic origin, being odd under time reversal, and coupling to static magnetic field gradients carrying a non-zero curl. However, $\bm{\tau}$ is also odd under a parity inversion~\cite{zeldovich1957}, just like an electric dipole, so that a system with toroidal polarisation can support linear magneto-electric coupling~\cite{spaldin2008toroidal}, and thus permits the electric manipulation of magnetic moments.

Single molecule toroics (SMTs) are a class of inorganic paramagnetic complexes that display a ground state molecular toroidal moment as a consequence of intramolecular vortex configurations made up of strongly anisotropic magnetic dipoles~\cite{tang2006, luzon2008spin, soncini2008toroidal, chibotaru2008origin, murray2022single}. Lanthanide ions are ubiquitous building blocks for SMTs owing to their strong axial magnetic anisotropy and comparatively weak intramolecular coupling~\cite{ungur2014single, li2019recent, vignesh2021strategies}, both essential ingredients to achieve quantum states carrying a toroidal moment~\cite{soncini2008toroidal}. Not only are SMTs interesting platforms to probe fundamental molecular physics beyond a simple magnetic dipole picture~\cite{soncini2008toroidal, luzon2008spin, pavlyukh2020toroidal}, but also they have been implicated in practical applications for hyper-dense molecular memory~\cite{spaldin2008toroidal}, molecular spintronics devices~\cite{soncini2010molecular, crabtree2018toroidal, rao2020toroidal} and as toroidal qubits~\cite{hymas2022role}. While realisations of toroidal moments have been reported in low-dimensional magnets~\cite{vanaken2007}, in metamaterials~\cite{zheludev2010} and in photonics applications~\cite{zheludev2022}, to date toroidal moments in SMTs have been extensively theorised and computed, but never observed.

The doubly degenerate ground states of the prototypical SMT, Dy$_3$, host a pair of counter-rotating toroidal vortex spin textures (related by time-reversal symmetry) with vanishing net magnetic moments~\cite{chibotaru2008origin}. The time- and parity-odd nature of the toroidal moment implies that selective preparation of just one toroidal configuration requires, in an achiral system, the application of a magnetic field gradient~\cite{spaldin2008toroidal, soncini2025arxiv}. However, engineering such gradients at the molecular level is particularly challenging and, to date, reports of sizeable and controllable static field gradients on the nanometre scale are absent from the literature, although a promising route has been recently proposed based on the coupling between the toroidal moment and the sufficiently fast ramp of an electric field, as achievable e.g. via ultrafast laser pulses~\cite{soncini2025arxiv}. The use instead of an electronic current density generated by an STM tip perpendicular to the molecular plane has also been proposed~\cite{schnack2022toroidalbasics, pavlyukh2020toroidal}, poised to address one or few molecules at a time in a monolayer of molecules orderly grafted on a conducting surface.  

\begin{figure}[t]
	\centering
	\includegraphics[width=0.48\textwidth]{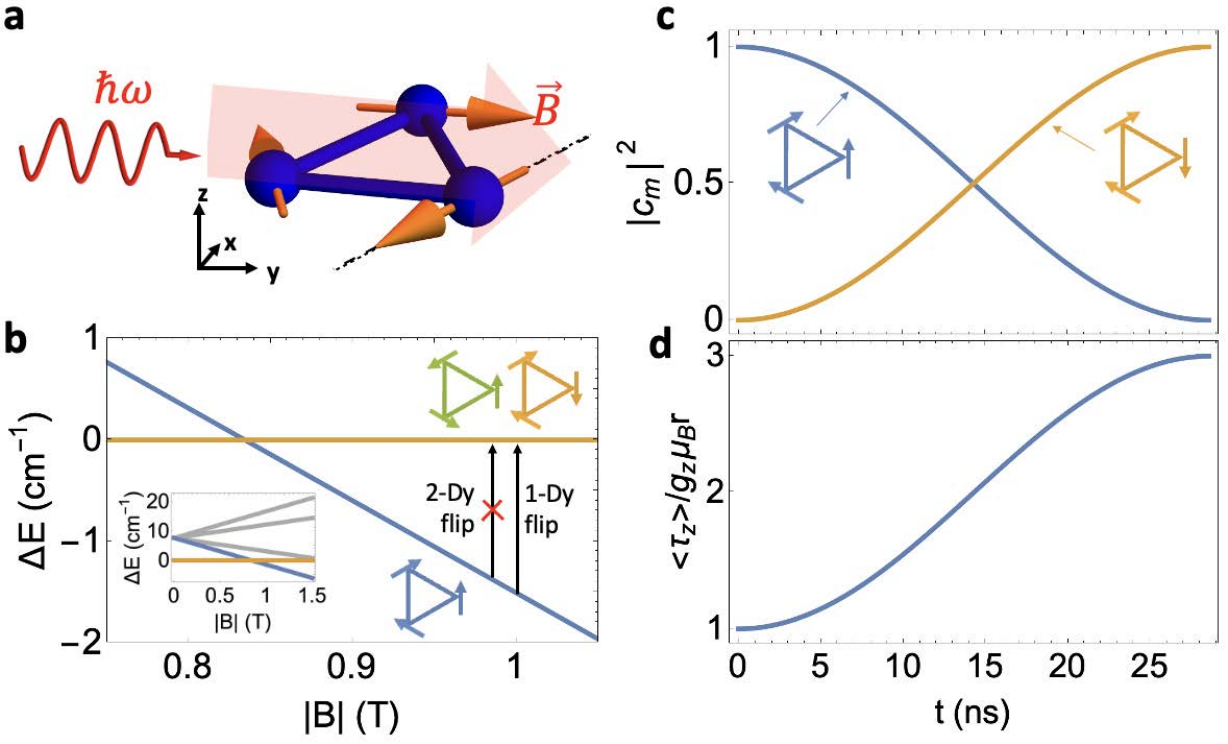}
	\caption{Prototypical preparation of a molecular toroidal moment. {\bf a} Schematic of the toroidal moment preparation experiment in the original Dy$_3$ SMT. Blue spheres represent Dy$^{3+}$ ions with principle magnetic axes (orange arrows) depicting their principal magnetic axes, arranged almost perfectly tangentially to the triangle vertices. A magnetic field is applied in the triangular plane, shown as a translucent red arrow, as well as a pulse of radiation with frequency $\omega$. {\bf b} Zeeman diagram of the states relevant to the EPR transition. The full Zeeman diagram is shown as an inset. {\bf c} Time evolution of the Dy$_3$ wavefunction with the system initialised in a static magnetic field $\mathbf{B} = \left(0, -1, 0 \right)$ and subject to a resonant pulse of radiation. {\bf d} Expectation value of the Dy$_3$ toroidal moment obtained from the wavefunction in {\bf c}.}
	\label{fig:1}
\end{figure}

Here, we identify a protocol for the controlled coherent manipulation of non-magnetic toroidal states in both Dy$_3$ and a family of its derivatives, MDy$_6$, in the context of a typical pulsed electron paramagnetic resonance (EPR) experiment.   A thorough computational and theoretical characterisation of the original Dy$_3$ SMT has been undertaken elsewhere~\cite{chibotaru2008origin} and revealed highly axial Dy$^{3+}$ magnetic moments directed almost perfectly tangential to the circumference of the triangular wheel (see Figure \ref{fig:1}a). We employ a non-collinear Ising model to describe the dynamics of the Dy$^{3+}$ pseudo spins $m_i = \pm 1$, akin to the model utilised in Ref.~\cite{chibotaru2008origin} which was able to convincingly capture the experimental powder magnetisation and magnetic susceptibility, and unequivocally demonstrate the toroidal character of the ground state doublet $\ket{\pm \tau} = \ket{\pm 1, \pm 1, \pm 1}$. When a finite field is applied in the plane of the Dy$_3$ triangle, the toroidal states become excited states of the system and the ground state is the singly degenerate magnetic state $\ket{-1, +1, +1}$. The Dy$^{3+}$ magnetic moments can be reoriented by a linearly-polarised resonant radiation propagating parallel to the triangular plane. Despite the degeneracy of the excited toroidal states, selective preparation of one over the other can be achieved with a single resonant $\pi$ pulse since the transition $\ket{-1, +1, +1} \rightarrow \ket{+ \tau} = \ket{+1, +1, +1}$ requires a single Dy flip (from a single resonant photon) whereas the transition $\ket{-1, +1, +1} \rightarrow \ket{- \tau} = \ket{-1, -1, -1}$ requires two (see Figure \ref{fig:1}b). Importantly, the $\ket{\pm \tau}$ degeneracy need not be removed via a magnetic field gradient at any point in this protocol.

Numerical simulations in Figure 1c and d justify this protocol. A $\pi$ pulse is implemented from the initialised $\ket{-1, +1, +1}$ Dy$_3$ magnetic state via integration of the time-dependent Schr\"odinger equation up to $t_p = \hbar \pi/g_x \mu_B \abs{B_{\perp}}$ accounting for radiation-magnetic dipole coupling between all Dy$^{3+}$ ions and a pulse of linearly polarised radiation propagating in the plane of the Dy$_3$ triangle. For the static field strength $\abs{B_{\parallel}} = 1$ T, a radiation frequency $\omega = 45$ GHz efficiently addresses the ground to first excited state gap in Dy$_3$. The pulse time chosen here is largely arbitrary, and has been selected to roughly coincide with recent pulsed EPR experiments on single-molecule magnets~\cite{takahashi2009coherent}. It may be attenuated or extended by varying the power of the applied radiation field $\abs{B_{\perp}}$. In Figure~\ref{fig:1}c we report the time-evolution of the squared amplitudes of the Dy$_3$ wavefunction which clearly demonstrates the selective preparation of just one toroidal state, hence maximising the toroidal moment in the Dy$_3$ triangle (Figure~\ref{fig:1}d).

While Dy$_3$ is not symmetric under inversion, it is worth noting that the unit cell of a Dy$_3$ crystal may show an inversion symmetry via the appropriate interchange of Dy$_3$ pairs. Such a unit cell would not be toroidally-polarisable since each pair of inversion symmetry related Dy$_3$ molecules would be prepared in opposing toroidal configurations, leading to a net zero toroidal moment. We re-emphasise this point when discussing MDy$_6$ below. In the absence of an inversion symmetry, however, the above protocol can result in a toroidal polarisation for the lifetime of the excited toroidal states that is detectable via the magneto-electric effect~\cite{spaldin2008toroidal}. Previous dynamical studies of MDy$_6$ complexes have approximated dysprosium spin-phonon relaxation constants between $10^{-2}$-$10^2$ s$^{-1}$ cm$^3$~\cite{vignesh2017ferrotoroidic, ashtree2021tuning} suggesting that a toroidal polarisation could be observable with a typical $\mu$-SQUID detection scheme.

\begin{figure}
	\centering
	\includegraphics[width=0.49\textwidth]{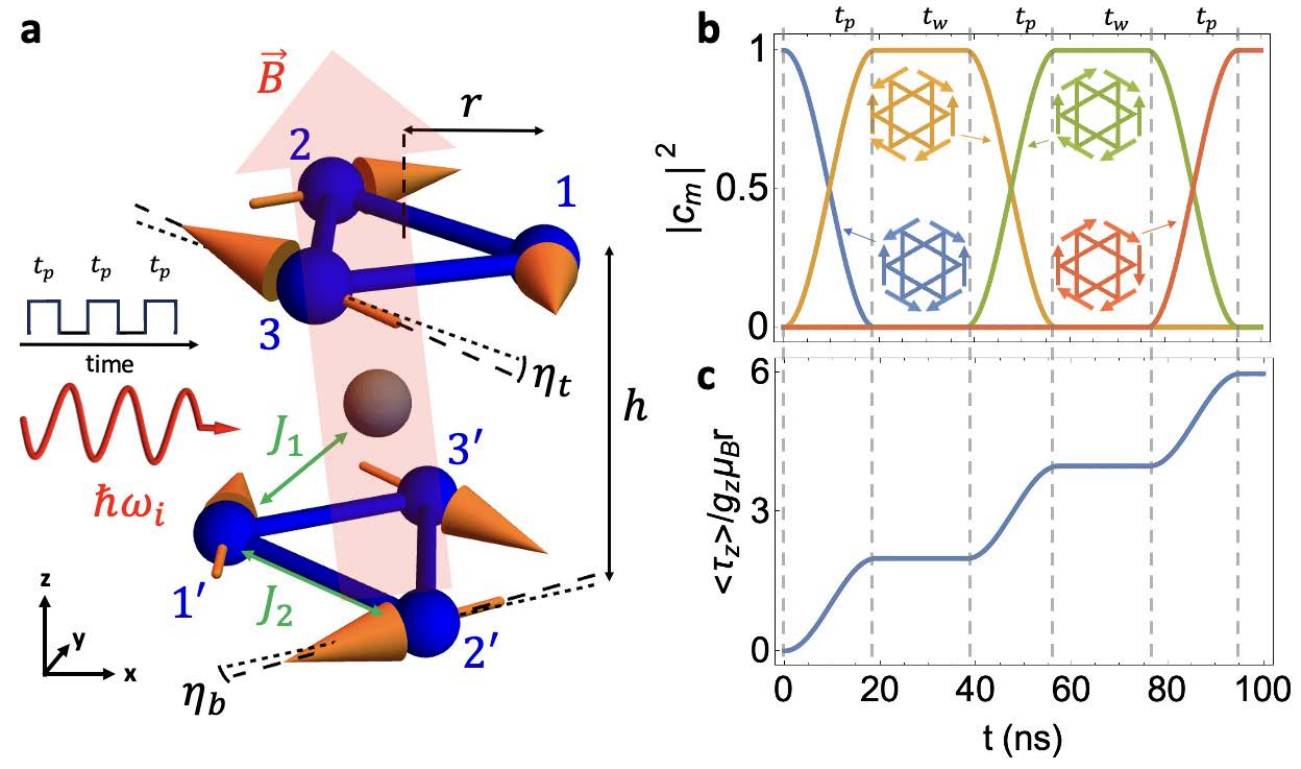}
	\caption{Selective preparation of a ferrotoroidic state with pulsed microwave radiation. {\bf a} Schematic of the AlDy$_6$ geometry where the Dy atoms are shown as blue spheres connected by blue pseudo-bonds. $h$ is the height of the molecule and the circumscribed radius of each SMT motif is $r$. Al is depicted as a grey sphere. In general, the lanthanide ions interact with the central transition metal ion via an exchange coupling $J_1$ and with nearest-neighbours in each SMT motif via $J_2$. The principal magnetic axes of the Dy ions are shown as orange arrows which are canted weakly out of the SMT triangular planes by the angles $\eta_t$ and $\eta_b$ for the top and bottom triangles, respectively. {\bf b} Time evolution of the AlDy$_6$ molecular wavefunction obtained from numerical integration of the time-dependent Schr\"odinger equation. The system is prepared in the AFT state $\ket{-\tau, +\tau}$ (blue curve) by the application of a static almost longitudinal field $\mathbf{B} = \abs{B_{\parallel}} \left( 0.1, 0, 1 \right)$ (shown as a red translucent arrow in {\bf a}) with $\abs{B_{\parallel}} = 1$ T. Three pulses of microwave radiation are applied to the system in pulses $t_p = \hbar \pi/ g_x \mu_B \abs{B_{\perp}}$ separated by the arbitrary wait times $t_w = t_p$ (grey dashed lines delineate the start and end of each pulse). After the completion of the three-pulse protocol, the FT state is prepared with $100 \%$ population (red curve). {\bf c} The time evolution of the molecular toroidal moment projected along the symmetry axis of the molecule $\langle \tau_z \rangle = \bra{\psi(t)} \tau_z \ket{\psi(t)}$.}
	\label{fig:2}
\end{figure}

The coupling of SMTs is imperative for both quantum computation based on toroidal qubits and for the development of ferrotoroidic metamaterials~\cite{zhang2022ferrotoroidic}. This has led to the synthesis of compounds containing multiple SMT moieties with a variety of coupling topologies~\cite{hewitt2010coupling, lin2012coupling, li2016planar, lin2017effect, kaemmerer2020inorganic, novitchi2012heterometallic, vignesh2017ferrotoroidic, vignesh2018slow}. We show how the EPR protocol above can be extended to manipulate the toroidal states in a class of non-trivially coupled SMTs (MLn$_6$) with two stacked triangular lanthanide moieties bridged by a central transition metal ion (Figure~\ref{fig:2}a). Tuning the molecular geometry of these compounds by modulating the central metal ion varies the ground state spin texture of the SMT, switching between con-rotating ferrotoroidic (FT) and counter-rotating antiferrotoroidic (AFT) configurations (typically with $\abs{\Delta E_{\text{AFT-FT}}} \sim 0.25$ cm$^{-1}$)~\cite{ashtree2021tuning}. When the AFT configuration is adopted, small deviations from an ideally tangential arrangement of the dysprosium magnetic axes (quantified by the canting angle $\eta$) results in a net magnetic dipole along the symmetry axis of the molecule. On the other hand, for the FT states, the inversion symmetry of the molecule enforces an exact cancellation of the net magnetic moment, no matter the value of $\eta$~\cite{vignesh2017ferrotoroidic}. While the selective preparation of AFT configurations is achievable by coupling to a uniform magnetic field along the symmetry axis of the molecule, selective coherent manipulation of the FT states is prohibited by the same limitations on molecular-scale magnetic field gradient generation as in Dy$_3$.

Computational and theoretical studies have previously revealed the strongly magnetically anisotropic nature of the constituent paramagnetic lanthanide ions of MDy$_6$ complexes~\cite{vignesh2017ferrotoroidic, ashtree2021tuning}. The principal magnetic axes adopt a tangential configuration to the vertices of the upper and lower triangular SMT moieties of each complex with an out-of-plane canting angle $\eta$ (Figure~\ref{fig:2}a). For real complexes, the canting angle is similar but not identical in the upper and lower triangles (designated $\eta_t$ and $\eta_b$ for top and bottom). The origin of this weak inversion symmetry breaking lies in the differential coordination of MeOH and NO$_3^-$ to the top and bottom SMT motifs~\cite{vignesh2017ferrotoroidic}. The weak symmetry lowering of MDy$_6$ in this way facilitates spectroscopic ferrotoroidic preparation in these complexes.

We consider ferrotoroidic state preparation in the AlDy$_6$ SMT where the diamagnetic Al$^{3+}$ bridging ion plays no role in the low-temperature spin dynamics. An AFT ground state $\ket{-\tau, +\tau}$ of AlDy$_6$ is prepared by the application of a (mostly) longitudinal magnetic field. A small in-plane component of the applied magnetic field ($10\%$ of the longitudinal component) lifts the degeneracies of high-lying single Dy-flip excited states so that individual Dy$^{3+}$ ions can be addressed via different frequencies of the radiation field. Contrary to the simpler case of Dy$_3$, to transition from the $\ket{-\tau, +\tau} \rightarrow \ket{+\tau, +\tau}$ now takes a minimum of three Dy flips, i.e. three resonant EPR pulses. Notably, if the molecule observes a perfect inversion symmetry, then the static magnetic field (which is time-odd but space-even) can never split in energy, configurations related by an inversion operation. Therefore, EPR-induced spin-flips in ions related by a spatial inversion will not be differentiated by our protocol and both ferrotoroidic states will be populated. 

Since an approximate inversion symmetry is predicted in real molecules, we proceed as above, now accounting for three resonant pulses of linearly polarised radiation propagating perpendicular to the symmetry axis of the molecule and assuming a discrepancy between the upper and lower triangle canting angles of $\eta_t - \eta_b = 1^{\circ}$. The pulse time $t_p = \hbar \pi/g_x \mu_B \abs{B_{\perp}}$ is constant for each pulse. On the other hand, the resonant frequencies of each pulse vary and are given by $\omega_1 = 115$ GHz, $\omega_2 = 43$ GHz and $\omega_3 = 101$ GHz for an applied field $\mathbf{B} = \abs{B_{\parallel}} \left( 0.1, 0, 1 \right)$ with $\abs{B_{\parallel}} = 1$ T. We show the time evolution of the AlDy$_6$ wavefunction in Figure~\ref{fig:2}b where we have inserted arbitrary wait times $t_w = t_p$ between each pulse for clarity. On completion of this pulse sequence, the molecule is selectively prepared in the $\ket{+\tau, +\tau}$ FT state, maximising the molecular toroidal moment of the complex without any application of a magnetic field gradient (Figure~\ref{fig:2}c).

\begin{figure}[b]
	\centering
	\includegraphics[width=0.49\textwidth]{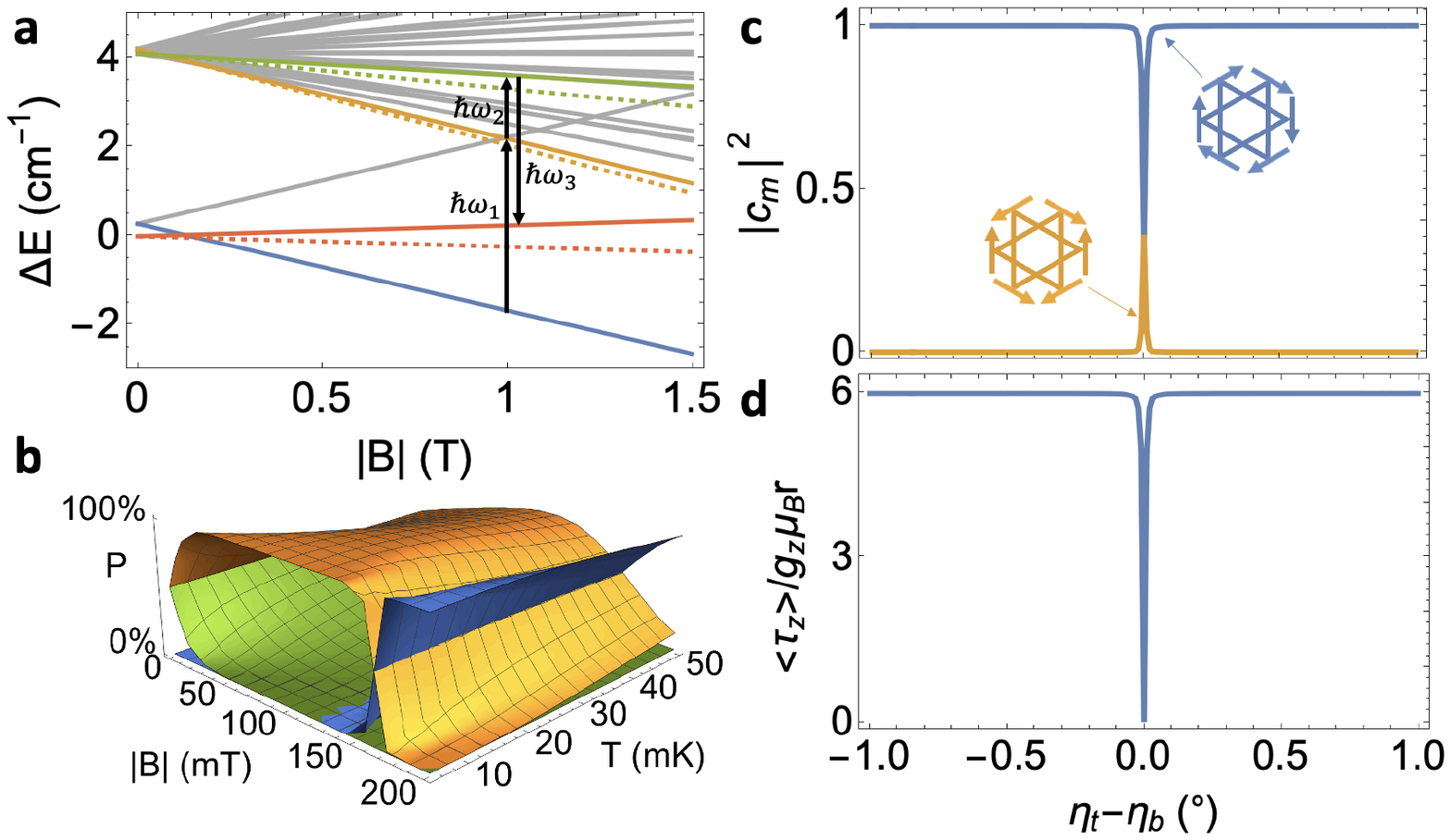}
	\caption{Fidelity of the ferrotoroidic preparation protocol. {\bf a} Zeeman diagram of AlDy$_6$ as function of the applied field $\mathbf{B} = \abs{B} \left(0.1, 0, 1 \right)$. The solid coloured levels indicate the Dy configurations under interrogation by the resonant pulsed radiation in Figure \ref{fig:2}b, dashed lines of the same colour depict related configurations which have EPR allowed transitions however lead to preparation of the time-reversed FT state. {\bf b} Boltzmann populations of the FT states $\ket{+\tau, +\tau}$ (yellow) and $\ket{-\tau, -\tau}$ (green) and the AFT state $\ket{-\tau,+\tau}$ (blue). {\bf c} Final squared amplitudes of the AlDy$_6$ wavefunction on completion of the three pulse protocol as a function of the differential canting angle of Dy$^{3+}$ principal magnetic axes in the upper and lower SMT moieties. {\bf d} Final expectation value of the AlDy$_6$ molecular toroidal moment as a function of the differential canting angle of Dy$^{3+}$ principal magnetic axes in the upper and lower SMT moieties.}
	\label{fig:3}
\end{figure}

The AlDy$_6$ state evolution for a fixed static field $\abs{B}$ is shown schematically in the Zeeman diagram of Figure~\ref{fig:3}a. Arrows denote the passage of population along a trajectory in the low energy Hilbert space of AlDy$_6$ (with colouring corresponding to Figure~\ref{fig:2}b) ending in complete ferrotoroidic population. Evidently, lifting the inversion symmetry AlDy$_6$ causes the FT states to acquire opposing magnetic moments (red solid and dashed levels in Figure~\ref{fig:3}a) suggesting a toroidal polarisation may be induced with a static magnetic field. In Figure~\ref{fig:3}b we plot the Boltzmann populations of the two FT states (in yellow and green) and one of the AFT states (blue).  Owing to the remarkably small splitting of the FT states in a static magnetic field, high fidelity selective population of just one FT state would require small applied fields and incredibly low temperatures $T < 10$ mK so as to render static FT preparation experimentally infeasible.

\begin{figure*}
	\centering
	\includegraphics[width=\textwidth]{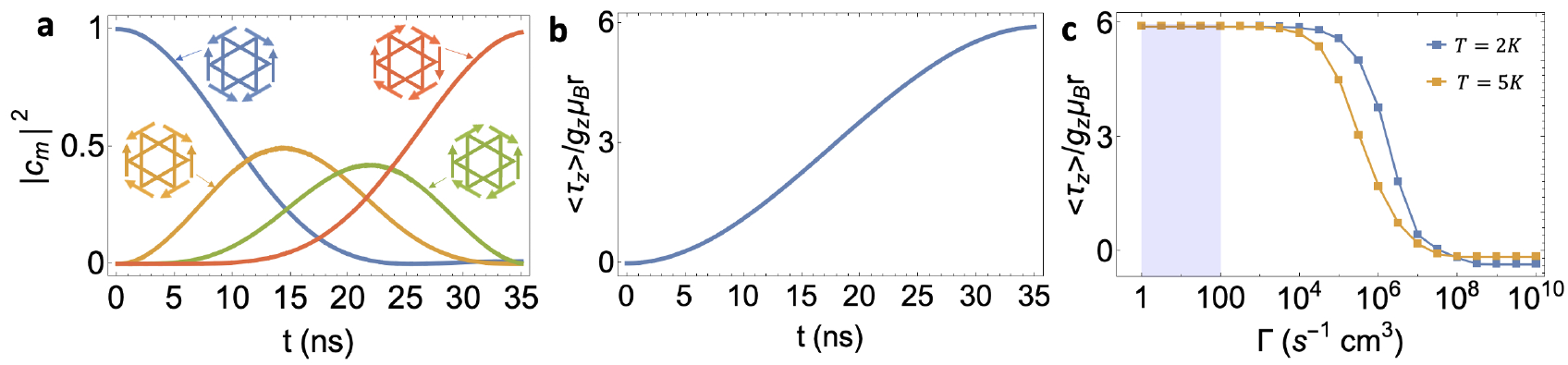}
	\caption{Coherent ferrotoroidic manipulation in an open quantum system. {\bf a} Time evolution of AlDy$_6$ populations when subjected to simultaneous microwave pulses. The system is prepared in the AFT state $\ket{-\tau, +\tau}$ (blue curve) by the application of a static almost longitudinal field $\mathbf{B} = \left(0.1 \ \text{T}, 0, 1 \ \text{T}\right)$. Three pulses of microwave radiation are applied simultaneously to the system over $35$ ns with amplitudes $B_{\perp}^{(1)}/B_{\perp}^{(2)} = B_{\perp}^{(3)}/B_{\perp}^{(2)} = 0.95$ and $B_{\perp}^{(2)} = 75$ mT. The FT state is prepared with $\sim 100 \%$ fidelity (red curve). {\bf b} The time evolution of the molecular toroidal moment expectation value projected along the symmetry axis of the molecule $\langle \tau_z \rangle$. {\bf c} The toroidal moment expectation value once the simultaneous three pulse protocol is concluded in the presence of incoherent spin-lattice relaxation with rate constant $\Gamma$ and bath temperature $T$. The shaded region indicates typical values of $\Gamma \sim 10^{-2} - 10^{2}$ s$^{-1}$ cm$^3$ for Dy$^{3+}$. Solid lines are added as a guide for the eyes.}
	\label{fig:4}
\end{figure*}

The dashed lines in Figure~\ref{fig:3}a are states that provide a three Dy-flip pathway from the ground AFT state (blue) to the time-reversed analogue of the target FT state (dashed red). When the inversion symmetry of the molecule is preserved, i.e. $\eta_t = \eta_b$, the solid and dashed lines are degenerate for all static fields. Both pathways are activated by the three pulse EPR sequence and it ceases to select between FT configurations. In Figure~\ref{fig:3}c and Figure~\ref{fig:3}d, we explore the final composition of the AlDy$_6$ wavefunction and the final toroidal moment, respectively, as a function of the differential canting angle in the upper and lower triangles. We find that symmetry breaking even by a fraction of a degree is significant to lift degeneracies between conjugate states so as to bias the coherent preparation of one ferrotoroidic state over another and result in a net toroidal polarisation.

Delivering multiple pulses simultaneously to the sample, as in nuclear double resonance spectroscopy~\cite{hartmann1962nuclear}, provides a strategy to mitigate incoherent relaxation of MDy$_6$ during ferrotoroidic preparation. We developed a minimal model for coherent population transfer between the ground AFT and excited FT state induced by three resonant pulses of microwave radiation delivered to the sample simultaneously (methods). By appropriately varying the amplitudes $B_{\perp}^{(k)}$ of the three pulses, an effective $\pi$ pulse can be achieved between the AFT and FT states. In Figure \ref{fig:4}a we plot the time-evolution of the AlDy$_6$ populations when all three microwave pulses are delivered simultaneously to the molecule. In contrast to the step-wise increase of the toroidal moment in Figure \ref{fig:2}c, under the simultaneous pulse protocol, the toroidal moment in Figure \ref{fig:4}b increases smoothly and monotonically to its maximum value as a consequence of only partial population of the single- and double-flipped Dy$^{3+}$ intermediate states (yellow and green curves in Figure \ref{fig:4}a).

Using a Redfield master equation to account for dissipative relaxation of AlDy$_6$ mid-pulse, we simulated a stochastic, memoryless thermal bath that induced spin-phonon mediated transitions between the energy eigenstates of AlDy$_6$. In Figure \ref{fig:4}c we report the toroidal moment expectation value on conclusion of the simultaneous pulse protocol, as a function of the single flip Dy$^{3+}$ rate constant $\Gamma$. For values of $\Gamma \sim 10^{-2} - 10^{2}$ s$^{-1}$ cm$^{3}$ in-line with current literature estimates~\cite{vignesh2017ferrotoroidic, ashtree2021tuning, staab2024intramolecular}, the simultaneous pulse protocol prepares the maximal toroidal moment of the complex. Incomplete toroidal polarisation results when the rate of incoherent relaxation is large compared to the timescale of the pulse sequence (in this case $\Gamma > 10^4$ s$^{-1}$ cm$^3$). The deleterious effect of spin-phonon coupling on ferrotoroidic preparation is two-fold. Not only does spin-phonon coupling redistribute population away from the intermediate and final excited states, but also $\Gamma$ accelerates the decay of the radiation-induced coherences (Supplementary Note 2). The onset of this strong dephasing regime is pushed to larger $\Gamma$ when decreasing temperature within the domain of existing EPR set-ups.

Additional paramagnetic species in the MDy$_6$ cluster lead to nuances in the experimental protocol. In Supplementary Note 3 we considered the more complicated situation of CrDy$_6$ under a similar three-pulse EPR experiment. For specific values of the applied static field, situations can arise where the energy gaps between same-Dy-configuration Cr$^{3+}$ spin states accidentally match the energy gaps targeted for selective ferrotoroidic preparation. Owing to the much larger (more than tenfold) transition magnetic dipole moment between Cr$^{3+}$ states, the Cr$^{3+}$ transitions are much stronger and impede FT preparation. Furthermore, the pulse times $t_p$ for resonant transitions between single-flipped dysprosium configurations are, in general, no longer identical. To achieve a population inversion between two single-flip dysprosium configurations $\mathbf{m}_i$ and $\mathbf{m}_f$, requires a pulse time $t_p = \hbar \pi/\abs{\bra{\phi_{f,\mathbf{m}_f}}\ket{\phi_{i,\mathbf{m}_i}}} g_x \mu_B \abs{B_{\perp}}$ where $\ket{\phi_{i,\mathbf{m}_i}}$ and $\ket{\phi_{f,\mathbf{m}_f}}$ are the initial and final eigenstates of the Cr$^{3+}$ wavefunction for each configuration. For moderately strong static fields, the Cr$^{3+}$ spin preferentially aligns to the field typically resulting in $\abs{\bra{\phi_{f,\mathbf{m}_f}}\ket{\phi_{i,\mathbf{m}_i}}} \approx 1$ but for static fields comparable to the Cr-Dy exchange and intramolecular dipolar coupling, the Cr$^{3+}$ wavefunctions are not necessarily aligned and longer pulse times are required for a full inversion. Both of these issues can be overcome in the high static field limit. We simulated high-fidelity coherent toroidal polarisation in CrDy$_6$ using the same protocol as AlDy$_6$ however with a static field only three times larger and aligned along the same spatial direction.

We have proposed two EPR experiments that utilise the broken inversion symmetry of Dy$_3$ and MDy$_6$, to selectively prepare maximal toroidal moments in each complex without the need for a molecular-scale magnetic field gradient. In the case of MDy$_6$, complete coherent population transfer between the AFT ground state and one of the FT states is achieved with either a sequential or a simultaneous microwave pulse protocol. The subsequent toroidal polarisation would be detectable with available $\mu$-SQUID set-ups via the magneto-electric effect. Our protocol provides an experimentally viable route to directly measure and coherently manipulate molecular toroidal states for the first time. A real implementation, made possible by our proposal, would establish molecular toroidal moments as a practical quantum degree of freedom and decisively advance its prospects for competitive next-generation qubit architectures and molecular spintronics devices. 

\bibliographystyle{ieeetr}

\cleardoublepage

\section*{Methods}

\subsection{Microwave-induced dynamics in an open system}

Th incoherent spin dynamics of the molecules studied in this work were obtained by deriving a Redfield master equation that accounted for the interaction of the paramagnetic spin system with a thermal bath of acoustic phonons. After tracing out the thermal bath, the equation of motion for the reduced density matrix elements $\rho_{mn}$ of the single-molecule toroic complex is
\begin{equation} \label{eq:m1}
\dot{\rho}_{mn} = -i \left[ H(t), \rho \right]_{mn} + \delta_{mn} \sum\limits_k W^{k \rightarrow m} \rho_k - \gamma_{mn} \rho_{mn}
\end{equation}
where $\rho_{mn}$ are populations ($m=n$) or coherences ($m \neq n$) between energy eigenstates of the molecule. The rates $W^{k \rightarrow m} \propto \Gamma$ and $\gamma_{mn} \propto \Gamma$ quantify the rate of Dy$^{3+}$ spin-flips owing to coupling with the phononic bath and are explicitly given in Supplementary Note 2. 

The coherent dynamics of the molecule are generated by $H(t) = H_0 + H_{\text{rad}}(t)$. The low energy spin dynamics for both Dy$_3$ and MDy$_6$ is faithfully captured by the effective Hamiltonian that models the lowest energy, maximal spin projection of each Dy$^{3+}$ ion as a non-collinear semi-classical spin. The dominant intramolecular mechanisms that establish the low energy physics of each complex are captured within the following Hamiltonian
\begin{equation} \label{eq:m2}
H_0 = H_{\text{ex}} + H_{\text{dip}} + H_{\text{Zee}}
\end{equation}
that accounts for intramolecular exchange, magnetic dipole coupling and a Zeeman interaction with a static magnetic field. $H_{\text{rad}}(t)$ semi-classically couples MDy$_6$ to a resonant, linearly polarised radiation is accomplished via 
\begin{equation} \label{eq:m3}
H_{\text{rad}}(t) = \sum\limits^{N}_i \mathbf{M}_i \cdot \mathbf{g}_i \cdot \mathbf{B}_{\text{rad}}(t)
\end{equation}
where $\mathbf{B}_{\text{rad}}(t)$ is the magnetic field of the resonant radiation. The radiation pulses take the form $\mathbf{B}_{\text{rad}}(t) = \sum^{3}_{k=1}  B^{(k)}_{\perp} \xi_k(t) \sin(\omega_k t) \mathbf{\hat{n}}$ where $\xi_k(t)$ is a square wave envelope that encapsulates the pulse, $\omega_k$ is the frequency of each pulse, $B^{(k)}_{\perp}$ is the magnitude of the field from each pulse and $\mathbf{\hat{n}}$ is perpendicular to the direction of radiation propagation.

To solve for the SMT dynamics we integrate Eq. (\ref{eq:m1}) numerically to obtain the time-evolution of the SMT reduced density matrix $\rho$. The molecular toroidal moment expectation value is given by $\langle \tau_z \rangle = \Tr\{\tau_z \rho\}$ where $\tau_z$ is the z-projection of the toroidal vector operator
\begin{equation} \label{eq:m4}
\boldsymbol{\tau} = g \mu_B \sum\limits_{i=1}^N \mathbf{r}_i \times \mathbf{M}_i
\end{equation}
where the sum extends over all paramagnetic ions. The coherent simulations were performed by numerically integrating Schrodinger's equation with the same Hamiltonian $H(t)$, equivalent to taking $\Gamma = 0$ in Eq. (\ref{eq:m1}).

\subsection{Analytical treatment of three simultaneous pulses}

The MDy$_6$ Hamiltonian in the absence of radiation coupling Eq. (\ref{eq:m2}) can be written as a spectral decomposition
\begin{equation} \label{eq:m5}
\begin{aligned}
H_0 &= \sum\limits_m \epsilon_m \ket{m} \bra{m}\\[2mm]
& \approx \epsilon_0 \ket{0} \bra{0} + \epsilon_1 \ket{1} \bra{1} + \epsilon_2 \ket{2} \bra{2} + \epsilon_3 \ket{3} \bra{3}
\end{aligned}
\end{equation}
where the ground $\ket{0}$, target $\ket{3}$ and one-Dy flip intermediate states $\ket{1}$ and $\ket{2}$ have been isolated in particular. The energies of these states follow the ordering $\epsilon_0 < \epsilon_3 < \epsilon_1 < \epsilon_2$, as per the set-up for ferrotoroidic preparation. These are the only states that display a non-trivial dynamics when coupled to the radiation field allowing us to neglect the dynamics of all other states in the MDy$_6$ Hilbert space. In this reduced basis, the radiation-matter coupling is
\begin{equation} \label{eq:m6}
\begin{aligned}
H_{\text{rad}}(t) &= \abs{B_{\perp}^{(1)}} \left( e^{-i \Omega_1 t} \ket{1} \bra{0} + e^{i \Omega_1 t} \ket{0} \bra{1} \right)\\[2mm]
& + \abs{B_{\perp}^{(2)}} \left( e^{-i \Omega_2 t} \ket{2} \bra{1} + e^{i \Omega_2 t} \ket{1} \bra{2} \right)\\[2mm]
& + \abs{B_{\perp}^{(3)}} \left( e^{i \Omega_3 t} \ket{3} \bra{2} + e^{-i \Omega_3 t} \ket{2} \bra{3} \right)
\end{aligned}
\end{equation}
where we have employed the rotating wave approximation. The Schr\"odinger equation is transformed to the rotating frame of the radiation via the unitary transformation $\ket{\psi(t)} = e^{-i H_0 t/\hbar} \ket{\psi_R(t)}$ to yield
\begin{equation} \label{eq:m7}
i \hbar \frac{d}{dt} \ket{\psi_R(t)} = e^{i H_0 t/\hbar} H_{\text{rad}}(t) e^{-i H_0 t/\hbar} \ket{\psi_R(t)} = H_R \ket{\psi_R(t)}.
\end{equation}
The rotating frame Hamiltonian $H_R$ can be evaluated explicitly by means of the Baker-Campbell-Hausdorff formula or via an equation of motion approach; we employ the latter. For example, with
\begin{equation} \label{eq:m8}
J_{10}(t) = \abs{B_{\perp}^{(1)}} e^{-i \Omega_1 t} e^{iH_0t/\hbar} \ket{1} \bra{0} e^{-iH_0t/\hbar}
\end{equation}
then
\begin{equation} \label{eq:m9}
\begin{aligned}
\frac{d J_{10}(t)}{dt} &= -i \left( \Omega_1 - \frac{\epsilon_1 - \epsilon_0}{\hbar} \right) J_{10}(t)\\[2mm]
\implies J_{10}(t) &= \abs{B_{\perp}^{(1)}} e^{-i \left(\Omega_1 - \omega_{10} \right) t} \ket{1} \bra{0}
\end{aligned}
\end{equation}
where $\omega_{10} = (\epsilon_1 - \epsilon_0)/\hbar$. All other terms on the RHS of Eq. (\ref{eq:m7}) are computed similarly leading to
\begin{equation} \label{eq:m10}
\begin{aligned}
H_R &= \abs{B_{\perp}^{(1)}} \left( \ket{1}\bra{0} + \ket{0} \bra{1} \right) + \abs{B_{\perp}^{(2)}} \left( \ket{2}\bra{1} + \ket{1} \bra{2} \right)\\[2mm]
& + \abs{B_{\perp}^{(3)}} \left( \ket{3}\bra{2} + \ket{2} \bra{3} \right)
\end{aligned}
\end{equation}
where $\Omega_1 = \omega_{10}$, $\Omega_2 = \omega_{21}$, and $\Omega_3 = \omega_{32}$. With $\ket{\psi(0)} = \ket{0}$ and $\ket{\psi(t_p)} = \ket{3}$, the general solution $\ket{\psi(t)} = e^{-iH_0t/\hbar} e^{-iH_Rt/\hbar} \ket{\psi(0)}$ is used to efficiently find a combination of $B_{\perp}^{(1)}$, $B_{\perp}^{(2)}$, $B_{\perp}^{(3)}$ and $t_p$ satisfying the simultaneous pulse toroidal moment preparation protocol.

\section*{Data Availability}
The Mathematica notebook used for the simulations in this work is available from the authors upon reasonable request.

\section*{Acknowledgements}
A. S. acknowledges financial support via the grant P-DiSC BIRD2023-UNIPD, from the Department of Chemical Sciences of the University of Padova, and from the Australian Research Council (Discovery Project no. DP210103208 and FT180100519). In addition, A. S. acknowledge funding from the University of Padova and Monash University Joint Initiative in Re- search (2024 Seed Fund scheme) for the project Single Molecule Toroics for Quantum Computation and the CINECA award under the ISCRA initiative, Project Grant SMTQUAN: Single Molecule Toroics for Quantum Computation, for the availability of high-performance computing resources and support. K. H. acknowledges financial support from the Revolutionary Energy Storage Systems Future Science Platform.

\section*{Author Contributions}
Both authors contributed to this work equally.

\section*{Competing Interests}
The authors declare no competing interests.

\end{document}


\preprint{APS/123-QED}

\title{Supplementary Information: Preparation and coherent manipulation of toroidal moments in molecules}

\author{Kieran Hymas}
\affiliation{Commonwealth Scientific and Industrial Research Organisation (CSIRO), Clayton, Victoria 3168, Australia}

\author{Alessandro Soncini}%
\email{alessandro.soncini@unipd.it}
\affiliation{Department of Chemical Sciences, University of Padova, Via Marzolo 1, 35131 Padova, Italy}

\date{\today}

\maketitle


\onecolumngrid

\section{Supplementary Note 1: Theoretical model}

For both Dy$_3$ and MDy$_6$, the magnetic behaviour of the lanthanide ions is faithfully described, in the low temperature limit, by a pseudo-spin Ising model with anisotropic $g$ tensors for each ion. This simplification is due to the relatively pure $\ket{m_J = \pm 15/2}$ ground doublets of each Dy$^{\text{3+}}$ ion, which are typically separated from excited crystal field states by $\gtrsim 100$ cm$^{-1}$. In the main text, we have focused our study on Dy$_3$ and AlDy$_6$ whose spin dynamics are captured completely from this semi-classical perspective. Additionally, we explored toroidal state manipulation in CrDy$_6$ where the paramagnetic Cr$^{3+}$ ion can not be accurately described as a semi-classical spin. We note here that previous {\em ab initio} studies~\cite{vignesh2017ferrotoroidic} revealed Cr$^{3+}$ to behave as a purely isotropic $S_M=3/2$ with $g=2.0023$ in CrDy$_6$.

We use this section to explicitly define our effective molecular Hamiltonian $H_0$ for the MDy$_6$ complexes but note that exactly the same model can be implemented for Dy$_3$ with straightforward adjustments following~\cite{chibotaru2008origin}. The relevant magnetic couplings between the constituent paramagnetic ions is captured by the Hamiltonian
\begin{equation} \label{eq:s1.1}
H_0 = H_{\text{ex}} + H_{\text{dip}} + H_{\text{Zee}}
\end{equation}
which accounts for intramolecular exchange, magnetic dipole coupling and a Zeeman interaction with a static magnetic field.

The exchange coupling Hamiltonian
\begin{equation} \label{eq:s1.2}
H_{\text{ex}} = -J_1 \sum\limits^{\text{Dy ions}}_{i} \mathbf{S}_i \cdot \mathbf{S}_M -J_2 \sum\limits^{\text{Dy ions}}_{i} \mathbf{S}_i \cdot \mathbf{S}_{i+1}
\end{equation}
accounts in the first term for an exchange coupling $J_1$ between the central metal ion and the six surrounding Dy ions and in the second term for the intratriangle Dy-Dy coupling $J_2$, where periodic boundary conditions are implied in the summand. To account for intramolecular magnetic dipole coupling we employ the Hamiltonian~\cite{hymas2022role}
\begin{equation} \label{eq:s1.3}
H_{\text{dip}} = \frac{\mu_0}{4 \pi} \sum\limits^{\text{All ions}}_{i \neq j} \left( \frac{\mathbf{M}_i \cdot \mathbf{M}_j}{\abs{\mathbf{R}_{ij}}^3} - 3 \frac{\left( \mathbf{M}_i \cdot \mathbf{R}_{ij} \right) \left( \mathbf{M}_j \cdot \mathbf{R}_{ij} \right)}{\abs{\mathbf{R}_{ij}}^3} \right)
\end{equation}
where $\mu_0$ is the permeability of free space, $\mathbf{M}_i$ is the magnetic moment of ion $i$ and $\mathbf{R}_{ij} = \mathbf{r}_i - \mathbf{r}_j$ is the displacement vector between ions $i$ and $j$. Lastly, we include a Zeeman interaction between the paramagnetic ions and a static external magnetic field $\mathbf{B}$ via the Hamiltonian
\begin{equation} \label{eq:s1.4}
H_{\text{Zee}} = \sum\limits^{\text{All ions}}_i \mathbf{M}_i \cdot \mathbf{B}.
\end{equation}

\cleardoublepage

\section*{Supplementary Note 2: Effect of decoherence on SMT coherent manipulation}

To model the time evolution of the quantum system in the presence of a thermal environment, we employed a Redfield equation~\cite{leuenberger2000spin, gatteschi2006molecular, vignesh2017ferrotoroidic, ashtree2021tuning}
\begin{equation} \label{eq:s3.1}
\dot{\rho}_{mn} = -i \left[ H(t), \rho \right]_{mn} + \delta_{mn} \sum\limits_k W^{k \rightarrow m} \rho_k - \gamma_{mn} \rho_{mn}
\end{equation}
for the reduced density matrix $\rho$ of the SMT quantum system. In Eq. (\ref{eq:s3.1}), $H(t) = H_0 + H_{\text{rad}}(t)$, $\delta_{mn}$ is the Kronecker delta function, $\gamma_{mn} = \frac{1}{2} \sum_k W^{m \rightarrow k} + W^{n \rightarrow k}$ and $W^{k \rightarrow m}$ corresponds to incoherent phonon-induced spin transitions between the energy eigenstates of $H_0$ indexed by $k$ and $m$. We model the phonon bath with a Debye model, leading to the phonon-induced spin relaxation rates
\begin{equation} \label{eq:s3.2}
W^{k \rightarrow m} = \Gamma \frac{\Delta_{mk}^3}{\exp(\Delta_{mk}/k_B T) - 1} \delta^{\text{1-flip}}_{mk}.
\end{equation}
In Eq. (\ref{eq:s3.2}), $\Gamma$ is the Dy$^{3+}$-flip rate constant, $\Delta_{mk} = \epsilon_m - \epsilon_k$ are energy gaps between the energy eigenstates of Eq. (\ref{eq:s1.1}), $k_B$ is Boltzmann's constant and $T$ is temperature. We consider phonon-induced transitions that flip one Dy$^{3+}$ at a time since these are likely to be the dominant contributors to the dissipative dynamics of the system. Accordingly, the term $\delta^{\text{1-flip}}_{mk}$ is equal to one if the eigenstates indexed by $m$ and $k$ can be connected by a single Dy$^{3+}$ flip, and is equal to zero otherwise. Typical values for single Dy$^{3+}$ flip rates are on the order $\Gamma \sim 10^{-2} - 10^2$ s$^{-1}$ cm$^{3}$~\cite{vignesh2017ferrotoroidic, ashtree2021tuning}.

In Figures~S1-3 we compare the time evolution of relevant populations and coherences of AlDy$_6$ during the simultaneous three pulse protocol with weak, intermediate and strong $\Gamma$. When $\Gamma = 10^2$ s$^{-1}$ cm$^3$ is weak, one of the two ferrotoroidic configurations (red) is prepared with high-fidelity owing to the sequential and maximal activation of the single- and double-Dy$^{3+}$ flipped coherences. In the intermediate regime, $\Gamma = 10^5$ s$^{-1}$ cm$^3$ suppresses population transfer by weakly dephasing the coherences of $\rho$. While a ferrotoroidic polarisation is achieved by the end of the pulse sequence, the protocol no longer operates at $100\%$ fidelity. When $\Gamma = 10^8$ s$^{-1}$ cm$^3$, dephasing dominates the dynamics quenching all coherent population transfer. Instead, population is redistributed to the first few low-lying states of the complex, partially populating both ferrotoroidic configurations.

\begin{figure}[h]
	\centering
	\includegraphics[width=0.75\textwidth]{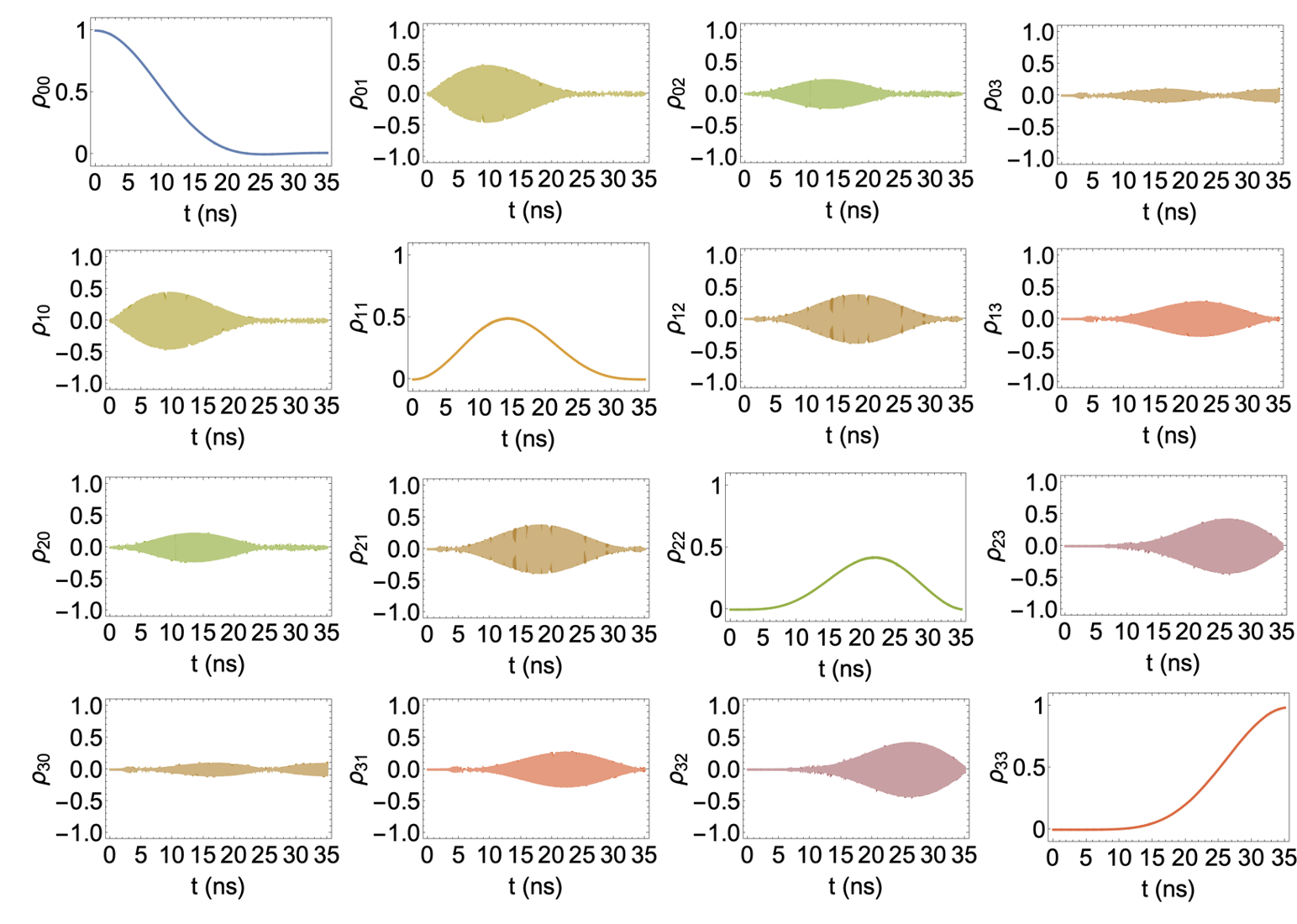}
	\caption{Time evolution of the populations and coherences of the AlDy$_6$ reduced density matrix under the simultaneous pump protocol described in the main text at $T=2$ K and $\Gamma = 10^2$ s$^{-1}$ cm$^3$. Only the populations and coherences coupled by the radiation field are shown.}
	\label{fig:s1}
\end{figure}

\begin{figure}[h]
	\centering
	\includegraphics[width=0.75\textwidth]{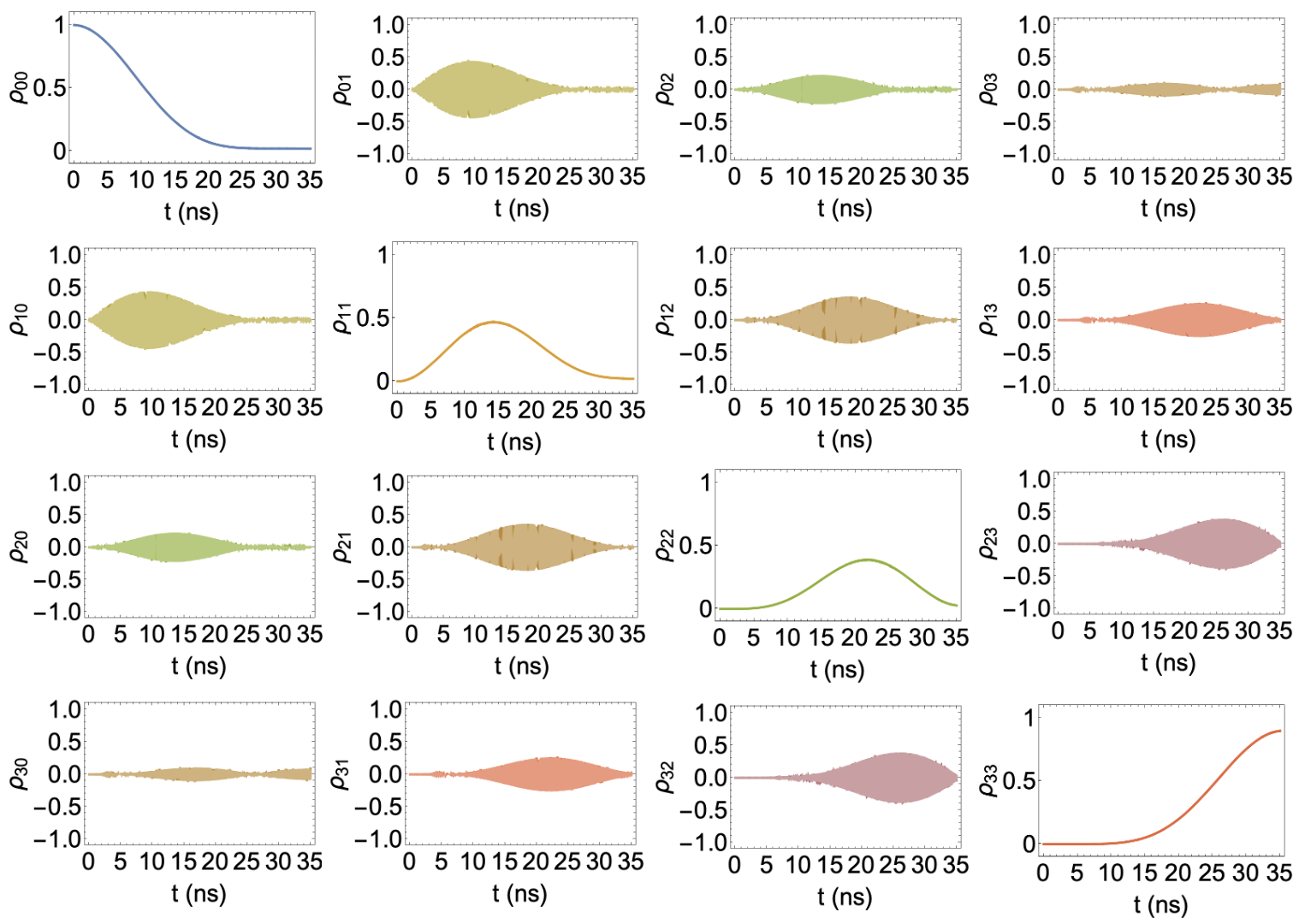}
	\caption{Time evolution of the populations and coherences of the AlDy$_6$ reduced density matrix under the simultaneous pump protocol described in the main text at $T=2$ K and $\Gamma = 10^5$ s$^{-1}$ cm$^3$. Only the populations and coherences coupled by the radiation field are shown.}
	\label{fig:s2}
\end{figure}

\begin{figure}[h]
	\centering
	\includegraphics[width=0.75\textwidth]{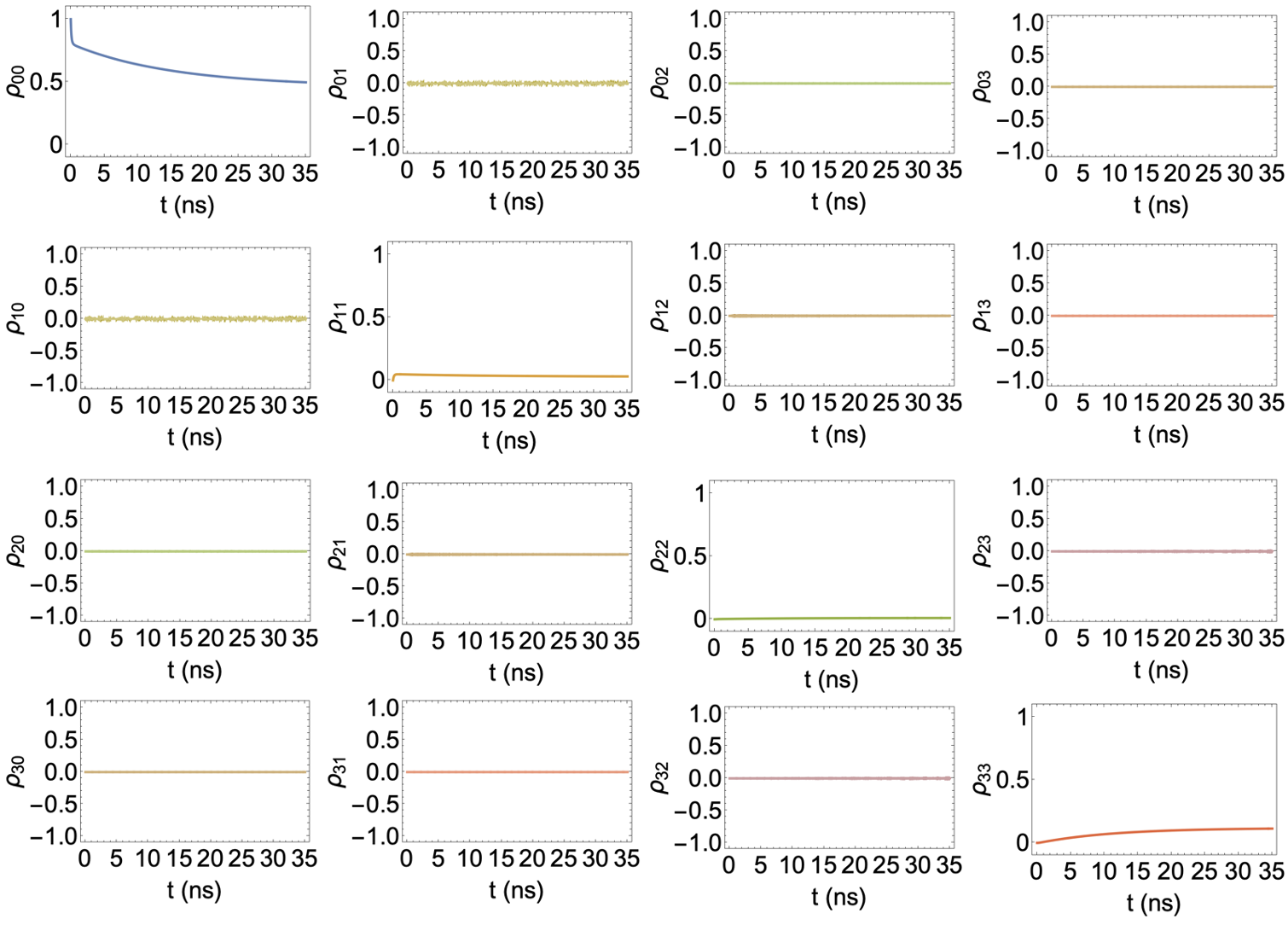}
	\caption{Time evolution of the populations and coherences of the AlDy$_6$ reduced density matrix under the simultaneous pump protocol described in the main text at $T=2$ K and $\Gamma = 10^8$ s$^{-1}$ cm$^3$. Only the populations and coherences coupled by the radiation field are shown.}
	\label{fig:s3}
\end{figure}

\cleardoublepage

\section*{Supplementary Note 3: Ferrotoroidic preparation in $\text{CrDy}_6$}

\begin{figure}[h]
	\centering
	\includegraphics[width=0.5\textwidth]{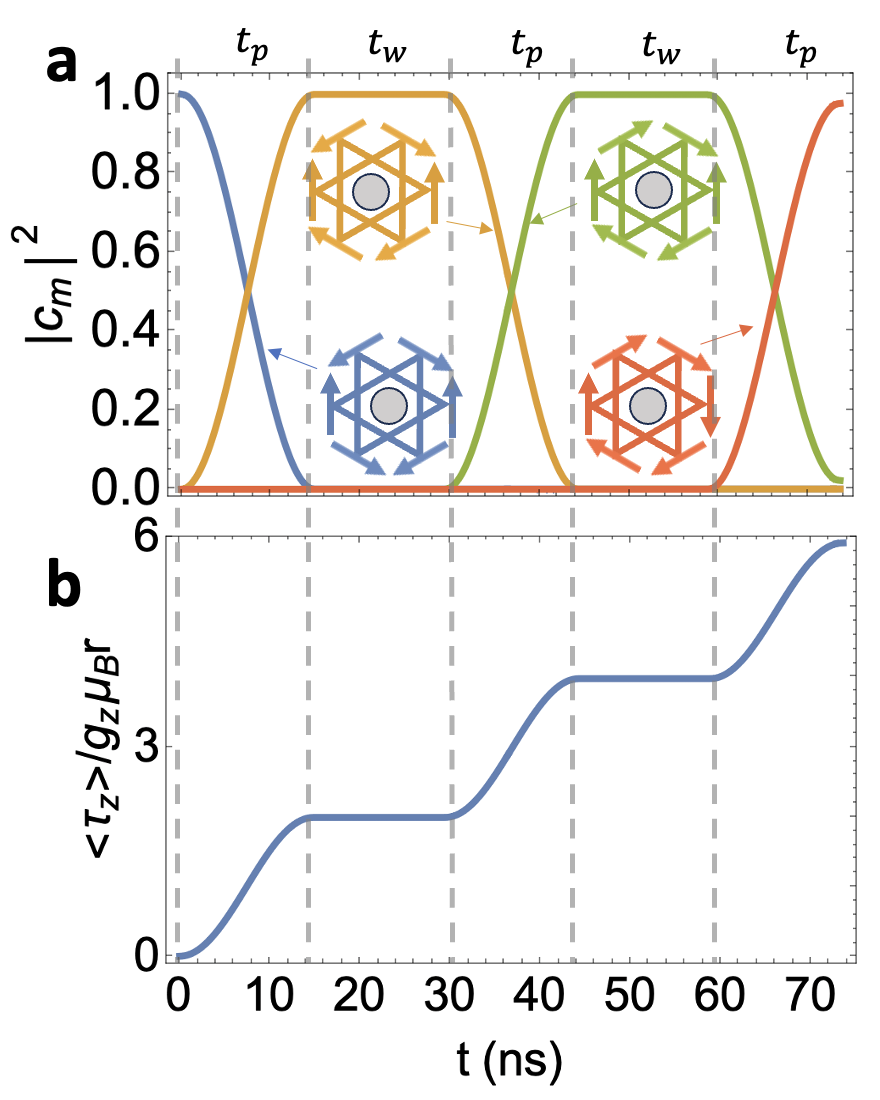}
	\caption{{\bf a} Time evolution of the CrDy$_6$ molecular wavefunction obtained from numerical integration of the time-dependent Schr\"odinger equation. The system is prepared in the AFT configuration $\ket{-\tau, +\tau}$ (blue curve) by the application of a static almost longitudinal field $\mathbf{B} = \abs{B_{\parallel}} \left(0.1, 0, 1 \right)$ with $\abs{B_{\parallel}} = 3$ T. Three pulses of microwave radiation are applied to the system in pulses of $t_p = 14$ ns separated by the arbitrary wait times $t_w = t_p$ (grey dashed lines delineate the start and end of each pulse). The frequency of each pulse is $\omega_1 = 183$ GHz, $\omega_2 = 150$ GHz and $\omega_3 = 56$ GHz. After the completion of the three-pulse protocol, the FT state is prepared with $100 \%$ population (red curve). {\bf b} The time evolution of the molecular toroidal moment projected along the symmetry axis of the molecule $\langle \tau_z \rangle = \bra{\psi(t)} \tau_z \ket{\psi(t)}$.}
\end{figure}
